\documentclass{appolb}
\usepackage{graphicx}
\usepackage[numbers,sort&compress]{natbib}

\newcommand{\be}{\begin{equation}}
\newcommand{\en}{\end{equation}}
\newcommand{\bea}{\begin{eqnarray}}
\newcommand{\ena}{\end{eqnarray}}
\newcommand{\dlangle}{\left\langle \kern-.17em \left\langle}
\newcommand{\drangle}{\right\rangle \kern-.17em \right\rangle}
\newcommand{\hbo}{\hbox to 1 true cm {\hfill } }
\newcommand{\tr}{\hbox{tr}}
\newcommand{\lb}{\langle \kern-.17em \langle} 
\newcommand{\rb}{\rangle \kern-.17em \rangle }

\newcommand{\DE}{\delta E}

\begin{document}
\title{From the density-of-states method to finite density quantum
  field theory %
\thanks{Excited QCD 2016.}%
}
\author{Kurt Langfeld$^a$, Biagio Lucini$^b$
\address{$^a$Department of Mathematical
Sciences,  University of Liverpool, \\
Liverpool, L69~7ZL, UK \\
Centre for Mathematical Sciences, Plymouth University,  \\
Plymouth, PL4 8AA, UK } 
\address{$^b$Department of Mathematics, Swansea University, 
Swansea SA2 8PP, UK}
}
\maketitle
\begin{abstract}
During the last 40 years, Monte Carlo calculations based upon
Importance Sampling have matured into the most widely employed method
for determinig first principle results in QCD. Nevertheless,
Importance Sampling leads to spectacular 
failures in situations in which certain rare configurations play a
non-secondary role as it is the case for Yang-Mills theories near a
first order phase transition or quantum field theories at finite
matter density when studied with the re-weighting method. The
density-of-states method in its LLR formulation has the potential to
solve such overlap or sign problems by means of an exponential error
suppression. We here introduce the LLR approach and its generalisation
to complex action systems. Applications include U(1), SU(2) and SU(3)
gauge theories as well as the Z3 spin model at finite densities and
heavy-dense QCD. 
\end{abstract}
  
\section{Introduction} 
Recently, Monte Carlo sampling methods for determining
density-of-states have seen a surge of interest. An
integral part of these novel density-of-state methods is a
re-weighting with the inverse density-of-states providing feedback
for an iterative refinement of this
quantity~\cite{Wang:2001ab}. Deriving the density-of-states in this
way, within chosen action intervals, allows us to obtain this
observable for regions of actions that conventional Importance
Sampling algorithms would never visit in practical simulation
times. For this reason, the density-of-state approach solves overlap
problems, which manifest when large tunnelling times, generally
growing exponentially with the size of the system, separate regions of
equally important statistical weight, henceforth causing an asymptotic
ergodicity problem in the latter algorithms. Methods based on
iterative refinements of the density-of-states fall into the class of
non-Markovian Random Walks. They extend outside the domain of
Importance Sampling the observation that a random walk in
configuration space is not plagued by exponentially large tunnelling
times~\cite{Berg:1992qua}.  In this paper, we focus on the Linear
Logarithmic Relaxation (LLR) algorithm~\cite{Langfeld:2012ah}, which 
is particularly suited for theories with continuous degrees of freedom,
and in particular for gauge theories. We here summarise the foundations of
the LLR method and its applications to gauge theories
(see~\cite{Langfeld:2015fua} and~\cite{Gattringer:2016kco}) and then
focus on recent successes of the LLR approach to quantum field
theories at finite densities~\cite{Langfeld:2014nta,Garron:2016noc}.

\section{The density-of-states approach }

\subsection{The Logarithmic Linear Relaxation (LLR) algorithm }

Our starting point is the partition function of an Euclidean quantum
field theory 
$$
Z(\beta) =\int  {\cal D} \phi \;  \e^{\beta  S [ \phi ]} \; . 
$$ 
The density-of-states $\rho (E)$ quantifies the amount of states if
the configurations are constrained to the action hyper-plane $E=
S[\phi]$. Its precise definition and relation to the partition function
are: 
\be 
\rho(E) =\int   {\cal D} \phi \;  \delta \Bigl( S[\phi]-E \Bigr) \; ,
\hbo 
Z(\beta) = \int dE \; \rho (E) \; \e ^{\beta E}  \; . 
\label{eq:1}
\en 
Our goal will be to obtain $\rho (E)$ with very high precision and to
directly calculate the partition function $Z$ by performing the
integral over $E$~\cite{Langfeld:2012ah}. To this aim, we divide the
action range into 
intervals of size $\DE$. If the action interval is small enough, we
can approximate the density-of-states by a Poisson-like distribution: 
$$ 
\rho (E) \propto \e ^{\alpha E} \, , \hbox to 1cm{\hfil for \hfil} 
E_k - \frac{\DE}{2}\leq E \leq E_k + \frac{\DE}{2} , \hbo 
\alpha_k = \frac{ d \ln \, \rho }{dE } \Big\vert_{E=E_k}. 
$$
Our strategy  is to calculate the LLR coefficients $\alpha _k$ and to
reconstruct $\rho (E)$. To this aim, we define the ``double-bracket''
Monte-Carlo expectation value with $a$ being an external parameter: 
\bea
\dlangle W[\phi] \drangle_{k} (a)
&=& \frac{1}{{\cal N}_k} \int {\cal D} \phi \; \theta
_{[E_k,\DE]}(S[\phi]) \; W[\phi] \; \,  \e^{-a S[\phi] } \; , 
\label{eq:2} \\
{\cal N}_k &=& \int {\cal D} \phi \; \theta _{[E_k,\DE]} \; \, 
\e^{-a S[\phi] } \; , 
\label{eq:3} 
\ena
where we have introduced the modified Heaviside function 
$$ 
\theta _{[E_k,\DE]} (S) \; = \; \left\{ \begin{array}{ll} 
1 & \hbox{for} \; \; \; E_k - \DE/2 \leq S \leq E_k + \DE /2 \\ 
0 & \hbox{otherwise . } \end{array} \right. 
$$
The key observation is that for $a = \alpha _k$, the probability
distribution for $\phi $ over the action interval becomes
flat since the re-weighting factor $\exp \{ - a S \}$ compensates the 
density-of-states $\exp \{ \alpha _k S\} $. Choosing $W[\phi ] =
S[\phi] - E_k$ as Litmus paper for flatness, we observe: 
\be 
 \dlangle S[\phi]  - E_k \drangle_{k} (a) \; = \; 0 \hbox to 2cm {\hfil
   for \hfil } a \; = \; \alpha _k \; = \;  \frac{ d \ln \, \rho }{dE
 } \Big\vert_{E=E_k}.  
\label{eq:4} 
\en 
This equation is at the heart of the LLR approach: it allows to
calculate the log-derivative $\alpha _k$ of the log derivative of
$\rho $ by solving the  stochastic non-linear equation 
$\dlangle S[\phi]  - E_k \drangle_{k} (a) \; = \; 0$ for $a$. 
We stress that the expectation values $\dlangle \ldots  \drangle $ are
accessible by standard Monte-Carlo simulations. 

\medskip 
A simple procedure to find the root of a function is the iterative
Newton-Raphson method
$$
a_{n+1}=a_{n} + \frac{ \dlangle S[\phi]  - E_k \drangle_{k} (a_n ) }{ 
\dlangle (S  - E_k)^2 \drangle_{k} } \approx 
a_n + \frac{12}{\DE^2} \, \dlangle S[\phi]  - E_k \drangle_{k} (a_n )
\; . 
$$
Note, however, that the statistical error from the Monte-Carlo
estimate for $ \dlangle S[\phi]  - E_k \drangle_{k} $ interferes with
the convergence of the Newton iteration. The solution to the
root-finding procedure for stochastic equations has been found by
Robinson and Monro. The iterative Robinson-Monro algorithm has the form
$$
a_{n+1}=a_{n}- c_{n}  \; \dlangle S[\phi]  - E_k \drangle_{k} (a_n ) 
, \qquad \mathrm{with} \ \sum_{n=0}^{\infty} c_{n}= \infty \
\mathrm{and} \ \sum_{n=0}^{\infty} c_{n}^{2} < \infty 
$$
Robinson-Monro proved that
$
\lim_{n \to \infty} a_n = \alpha _k 
$
with $a_n$ asymptotically normally distributed around $\alpha _k $. 
 To minimise the variance of the result one chooses
$$
c_n = \frac{12}{\DE ^2 \, (n+1)}  \; . 
$$
Once the LLR coefficients $a_k$ are obtained for the action range,
the density-of-states can be obtained by
\be 
\rho (E)=\prod_{i=1}^{k-1} e^{{a}(E_i) \frac{\DE}{2}}
\exp \Bigl( a(E_k)(E-E_k) \Bigr) \,  , \qquad E_k \leq E < E_{k+1} \; . 
\label{eq:5}
\en 

The LLR method has some remarkable features~\cite{Langfeld:2015fua}:
\begin{itemize}
  \item[(i)] Almost everywhere, we find that the LLR approximated
    result $\rho (E)$ (\ref{eq:5}) is related to the {\it exact}
    density-of-states by

$$
\rho(E) \;  \e^{\gamma_1 \, \DE ^2} \; \le \; \rho_{\mathrm{exact}}(E)
\; \le \; \rho(E) \;  \e^{\gamma_2 \, \DE ^2} \ , \ \gamma_1, \ \gamma_2
\ \mathrm{constants}.
   $$

The approach has {\it exponential error suppression}:
  the relative approximation error does not depend on the magnitude of
  $\rho$ despite $\rho $ might span thousands of orders of
  magnitude.

  \item[(ii)] The LLR approach can be generalised to calculate
    expectation value of arbitrary observables (rather than the
    partition function only). The systematic error is ${\cal O}(\DE ^2)
    $. See~\cite{Langfeld:2015fua} for details. 
\end{itemize}

\subsection{The generalised density-of-states}

Quantum Field Theories at finite matter densities (or more precisely,
at non-vanishing chemical potential $\mu $) are generically
plagued by the so-called {\it sign problem}. In this case, the
partition function, which features a complex action, can be written in
the general form as 
\be 
Z(\beta, \mu) = \int {\cal D} \phi \; \exp \Bigl\{ \beta \, S_R[\phi] + i
\, \mu \, S_I[\phi] \Bigr\} \; . 
\label{eq:6}
\en
The Gibbs factor looses the interpretation
as probability density and Importance Sampling is impossible.
An early attempt to circumvent the sign problem was to drop the phase
factor when generating the lattice configuration and to add the phase
factor to the observable:
$$
\langle A \rangle \; = \; \frac{ \langle A \, \exp \{ i\, \mu \,
  S_I[\phi] \} \rangle _\mathrm{PQ} }{ \langle  \exp \{ i\, \mu \,
  S_I[\phi] \} \rangle _\mathrm{PQ} } \; , \hbo 
Z_\mathrm{PQ} \; = \; \int {\cal D} \phi \; \exp \Bigl\{- \beta \,
S_R[\phi] \Bigr\} \; . 
$$
In this phase quenched approach, we encounter an overlap
problem. Following~\cite{Langfeld:2014nta,Gattringer:2016kco}, we
define the overlap between full and phase quenched theory by the ratio
of their partition functions, i.e.,  
\be
Q(\mu) \; = \; \frac{ Z(\mu) }{ Z_\mathrm{PQ} (\mu) } \; = \;
\left\langle \exp \{ i\, \mu \,  S_I[\phi] \} \right\rangle
_\mathrm{PQ} \; . 
\label{eq:7}
\en
Since the phase-quenched theory has a positive probabilistic measure, we
find by virtue of the triangular inequality that~\cite{Gattringer:2016kco} 
$$
\vert Q(\mu) \vert \leq \Bigl\langle \vert \exp \{ i\, \mu \,
S_I[\phi] \} \vert \Bigr\rangle _\mathrm{PQ} \; = \; 1 \; . 
$$
Note, however, that both theories generically differ in their free
energies $f$, leading to exceptionally small overlaps at large volumes
$V$:
$$
Q(\mu) \; = \; \exp \Bigl\{ - \, \Delta f \; V \Bigr\} \; ,
\hbo \Delta f \ge 0 \; . 
$$
Since the LLR method generically solves overlap problems, we extend 
the approach outlined in the previous subsection and
define the {\it generalised density-of-states} by
\be 
\rho _\beta (s) \; = \; N \; \int {\cal D} \phi \; \delta \Bigl( s \, - \, 
S_I[\phi](\mu) \, \Bigr) \;  \e ^{\beta \, S_R[\phi ](\mu )} \; ,
\label{eq:8}
\en
with $N$ an unknown normalisation factor independent of $s$. The
overlap then appears as the Fourier transform of the generalised
density: 
$$
Q(\mu ) \; = \; \frac{ \int ds \; \rho_\beta (s) \; \exp (is ) }{ \; 
\int ds \; \rho _\beta (s) } \; . 
$$
Note that the unknown normalisation $N$ has dropped out.

\section{Vacuum applications: U(1), SU(2) SU(3) Yang-Mills theories }

\begin{figure}[htb]
\centerline{%
  \includegraphics[width=6.2cm]{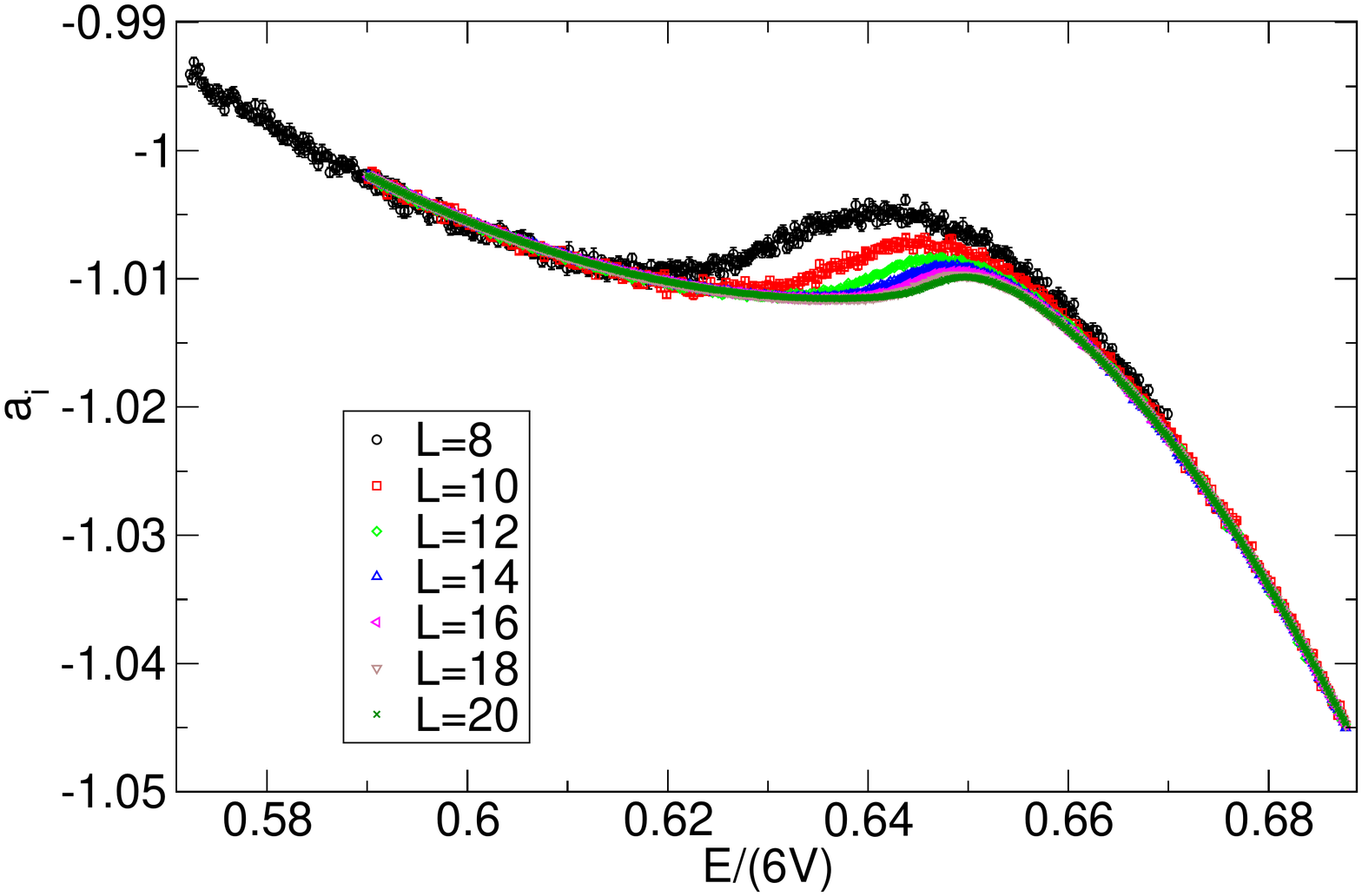} \hspace{0.1cm}
  \includegraphics[width=6.2cm]{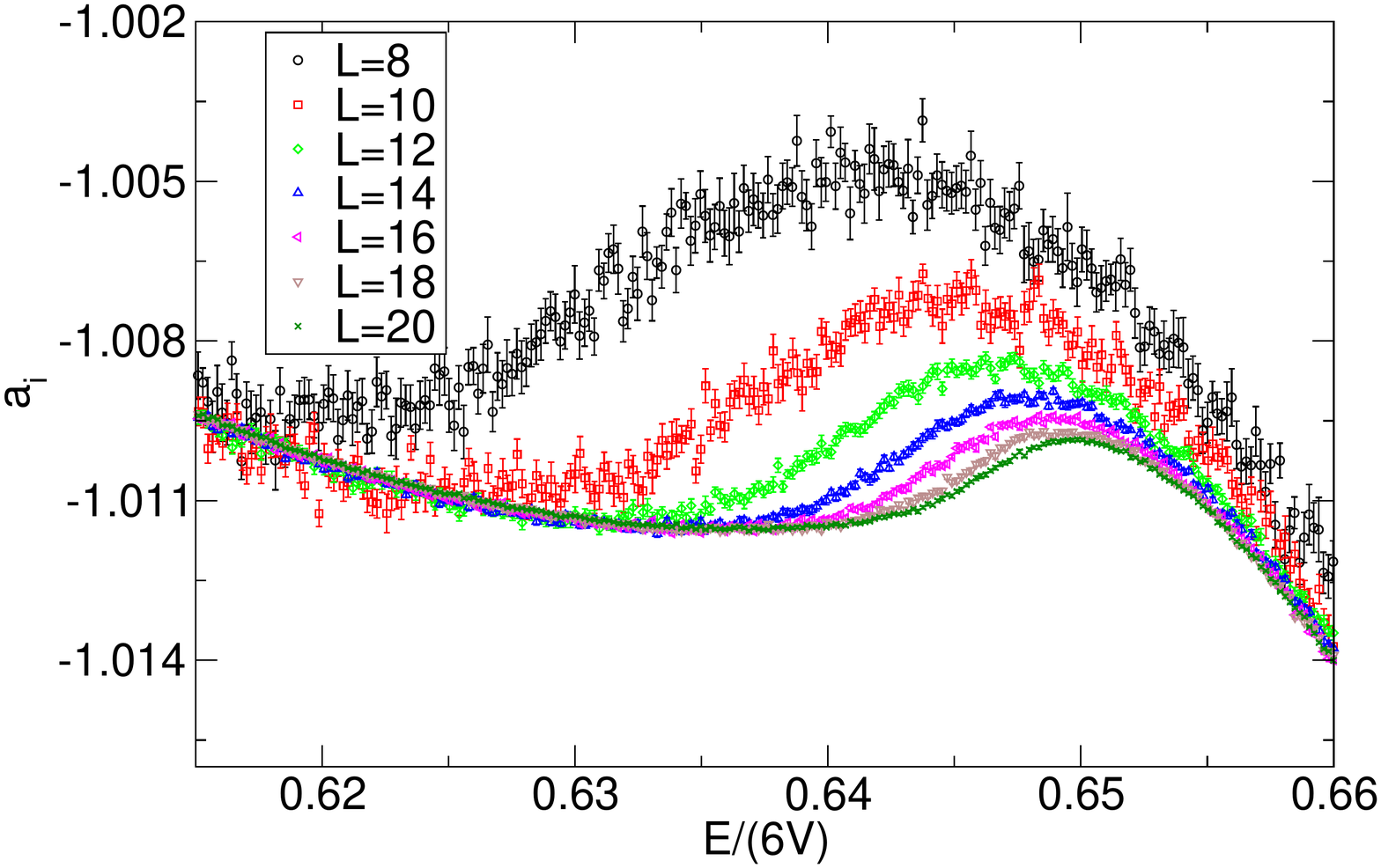}
}
\caption{LLR coefficient $a(E)$ (or $\alpha (E) = d \, \ln \rho /dE$)
  as a function of the action $E$ for a 
  U(1) gauge theory for several lattice volumes. }
\label{fig:1}
\end{figure}
In the following, we show the LLR method in action for pure gauge
theories in vacuum, i.e., without finite density matter and no-sign
problem. The degrees of freedom are the link fields
$U_\mu (x) \in $U(1), SU(2) or SU(3), and the action is given by 
$$
S \; = \;  \sum _{\nu>\mu, x} \frac{1}{N_c} \, \mathrm{Re} \,
\tr \{ U_\mu (x) U_\nu (x+\mu) U^\dagger _\mu (x+\nu) U^\dagger _\nu
(x) \, \} \; , 
$$
where $N_c$ is the number of colours ($N_c=1$ for U(1)), the 'Re' can
be omitted for SU(2), since the theory is real, and the trace 'tr' is
absent for U(1). The empty (perturbative) vacuum is attained with $S=
6V$ where $V$ is the volume, i.e., the number of lattice points. 

\begin{figure}[htb]
\centerline{%
  \includegraphics[width=6.2cm]{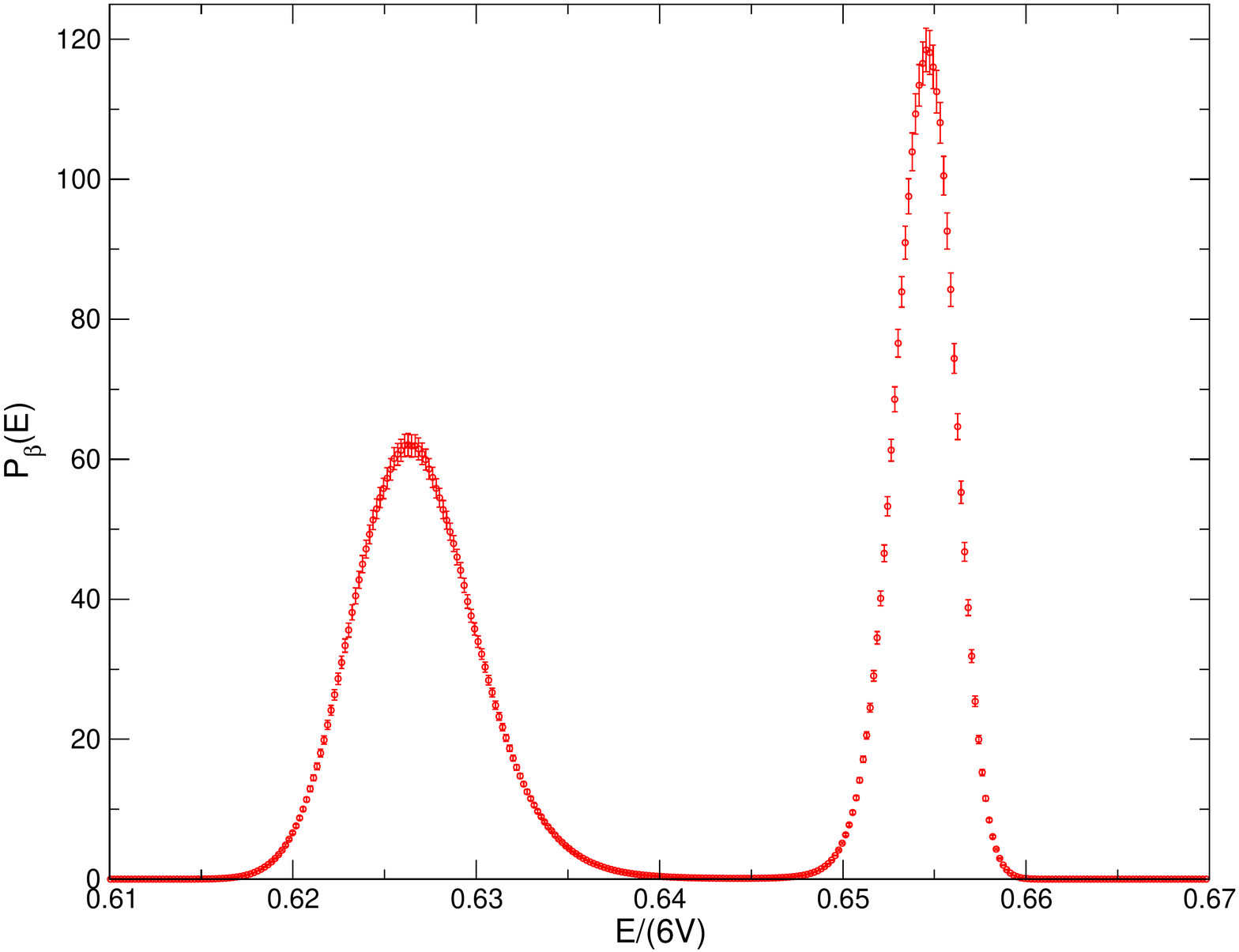} \hspace{0.1cm}
  \includegraphics[width=6.2cm]{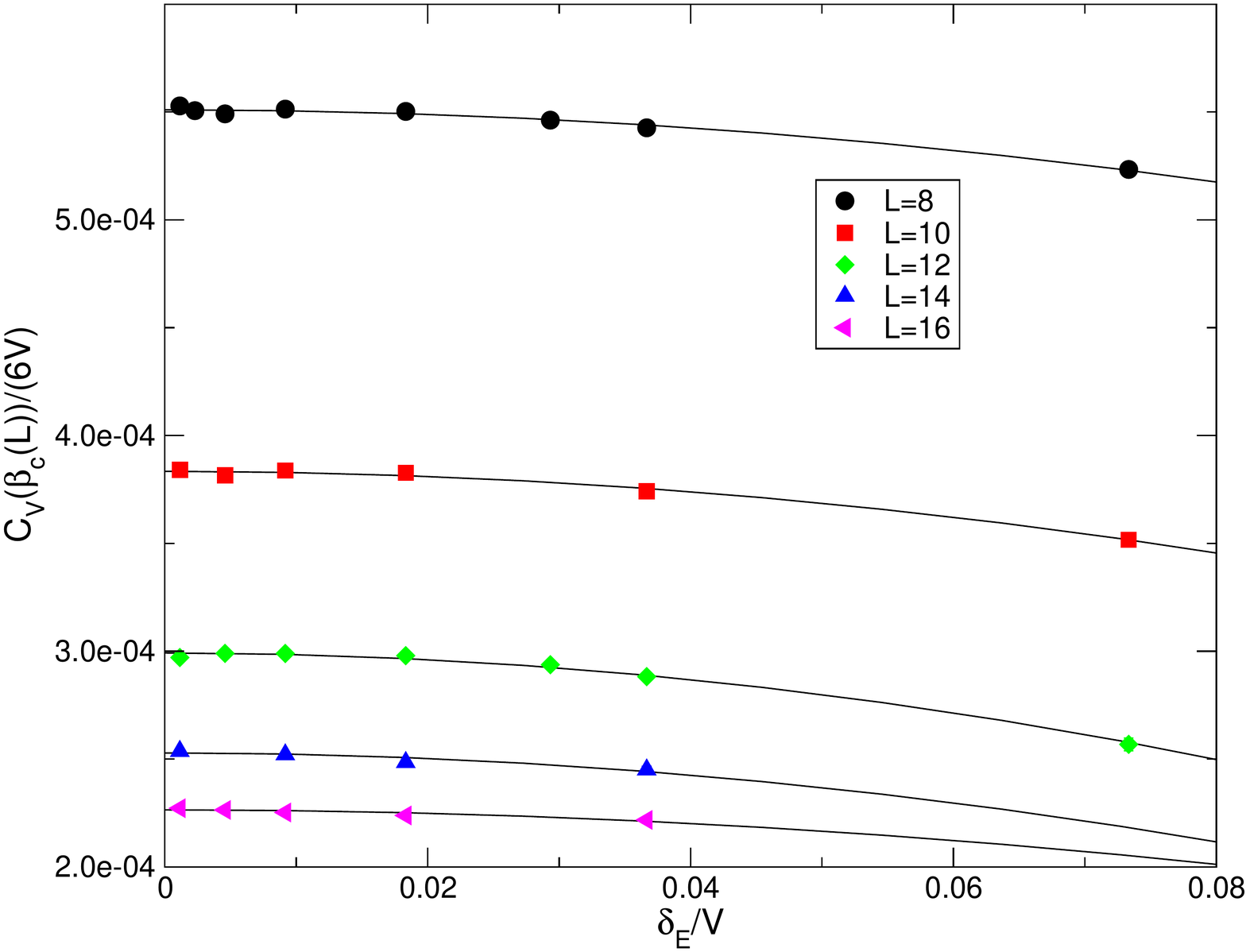}
}
\caption{Left: Probability distribution of the action for a U(1) gauge
  theory on a $20^4$ lattice at criticality. Right: Specific heat as
  function of the interval discretisation length $\DE$ (results
  from~\cite{Langfeld:2015fua}).   }
\label{fig:2}
\end{figure}
For each of these theories, the probabilistic measure rises from the
product of the density-of-states and the  Gibbs factor:
\be
P(E) \; = \; \rho (E) \, \exp \{ \beta E \} \; . 
\label{eq:10}
\en
\begin{figure}[htb]
\centerline{%
  \includegraphics[height=6.2cm]{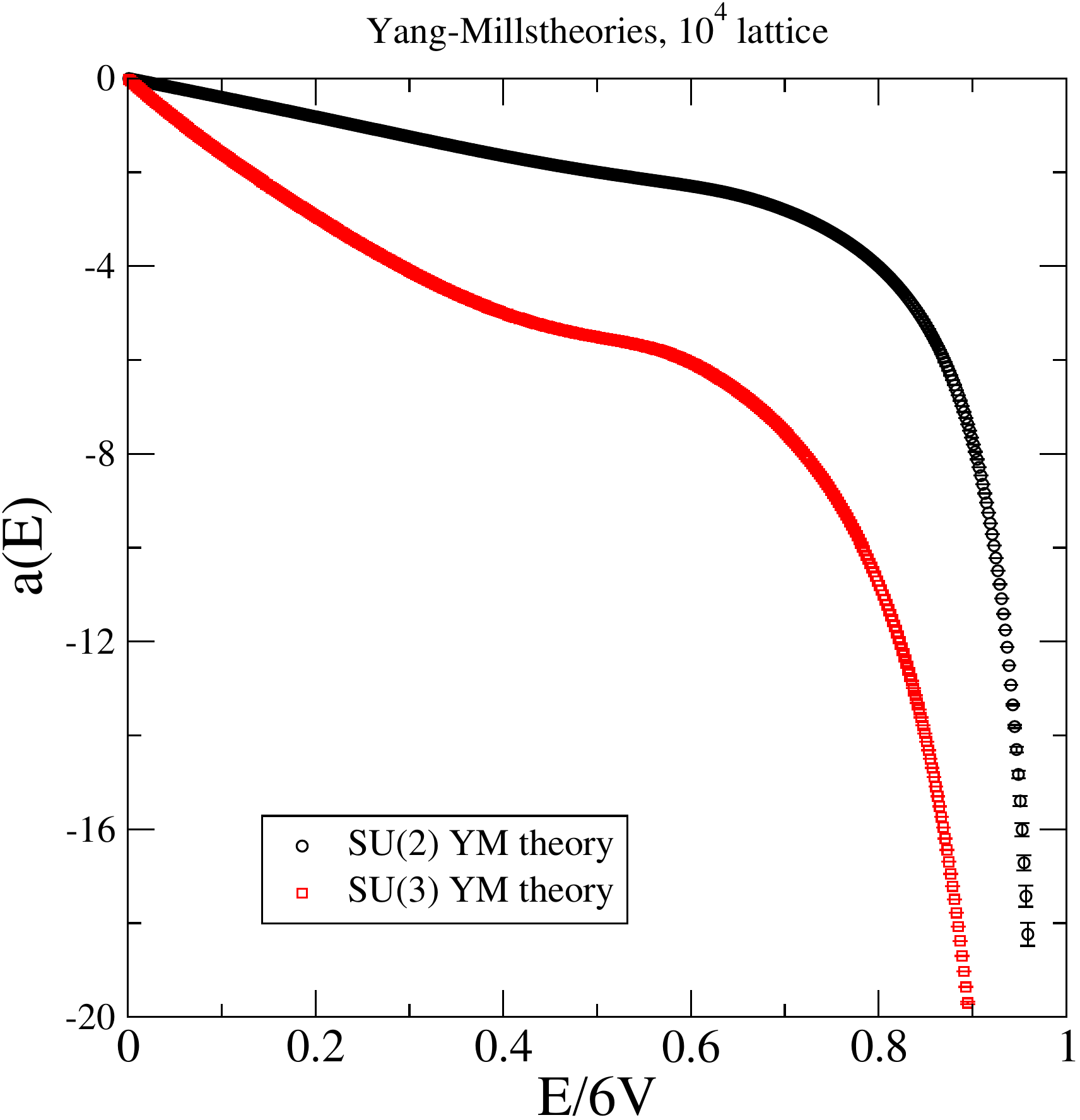} \hspace{0.1cm}
  \includegraphics[height=6.2cm]{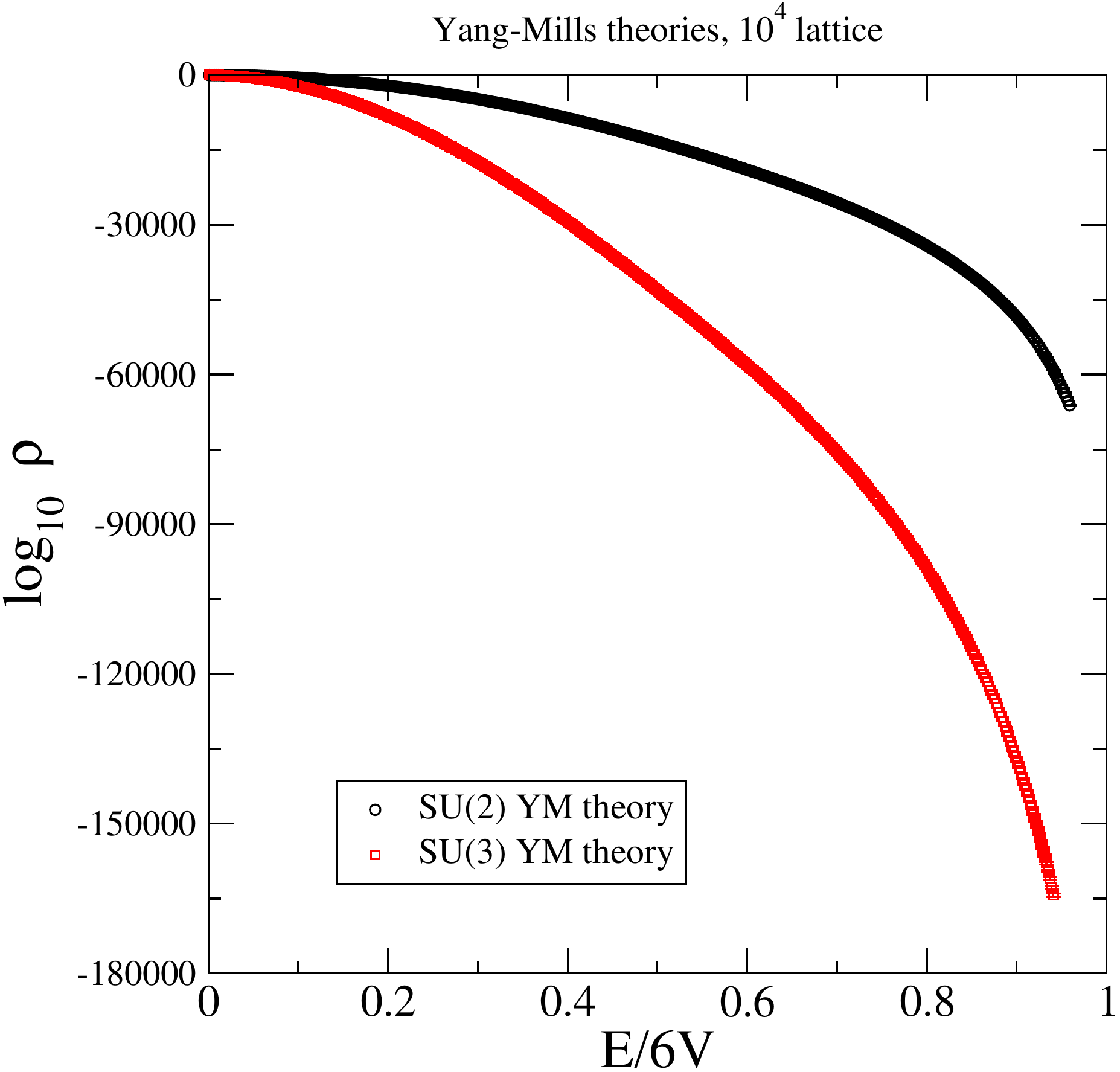}
}
\caption{Left: The LLR coefficient $a(E)$ as a function of the action
  $E$ on a $10^4$ lattice for a SU(2) and SU(3) Yang-Mills
  theory. Right: The corresponding density-of-states. Results
  from~\cite{Gattringer:2016kco}. } 
\label{fig:3}
\end{figure}
Generically, $\rho (E)$ monotonically decreases with $E$ while the
Gibbs factor exponentially increases. Thus, $P(E)$ settles for a sharp
maximum (away from first order criticality). The most likely value for the
action can be calculated from: 
\be 
\frac{dP(E)}{dE} \; = \;  \rho (E) \, \exp \{ \beta E \} \Bigl[
    \alpha (E) \; + \; \beta \Bigr] \; = \; 0 \;  
\label{eq:11}
\en 
Away from criticality, this equation most have just one solution for
each $\beta $. This observation can be used to calculate $\rho (E)$
for small $E$ using the strong coupling
expansion~\cite{Gattringer:2016kco}. For a first order phase
transition at $\beta = \beta _c$, $P(E)$ exhibits the typical
double-peak structure. Hence, $P(E)$ features two maxima and one
minimum meaning that (\ref{eq:11}) has three solutions. Let us
illustrate this for the U(1) gauge theory~\cite{Langfeld:2015fua}, for
which our
findings for the LLR coefficient are shown in figure~\ref{fig:1}. In
accordance with the theory~\cite{Langfeld:2015fua}, $a(E)$ becomes
volume independent at large volumes. In the region between $E/6V =
0.61$ and $E/6V = 0.66$, we observe a non-monotonic behaviour that
leads to three solutions of the equation (\ref{eq:11}) for a suitably
chosen $\beta $. Our result for $P(E)$ for a lattice as large as
$20^4$ is shown in figure~\ref{fig:2}. We stress that our result 
is obtained for an un-rivalled lattice and that we do not see a
significant critical slowing-down while increasing the lattice
size. A more detailed analysis of the volume scaling properties of our
algorithm is left to future work. Using the specific heat $C_v$, we
also studied the systematic errors induced by the finite action
interval size $\DE $. Figure~\ref{fig:2} shows $C_V$ as a function
of $\DE$ for several values $\DE$. Our numerical results suggest a
quadratic behaviour in $\DE $, which is in accordance with the
theory~\cite{Langfeld:2015fua}. 

\medskip
We finally show the results for the SU(2) and SU(3) Yang-Mills theory
on a $10^4$ lattice. Figure~\ref{fig:3} shows $a(E) = d \, \ln \rho /
dE$ as a function of the action $E$. Also shown is the leading order
analytical result at small $E$~\cite{Gattringer:2016kco}. In the same
figure, we also show the reconstructed density-of-states
(\ref{eq:5}). The error bars were obtained by a bootstrap analysis of
$40$ independent results for $a(E)$ for each $E$. Note the logarithmic
vertical axis: for SU(3), we obtain the density-of-states over
$150,000$ orders of magnitude with an almost constant statistical
error bar over the whole action range.

\section{Applications to finite density quantum field theories}

\subsection{The $Z_3$ theory as showcase }

\begin{figure}[htb]
\centerline{%
  \includegraphics[height=5.cm]{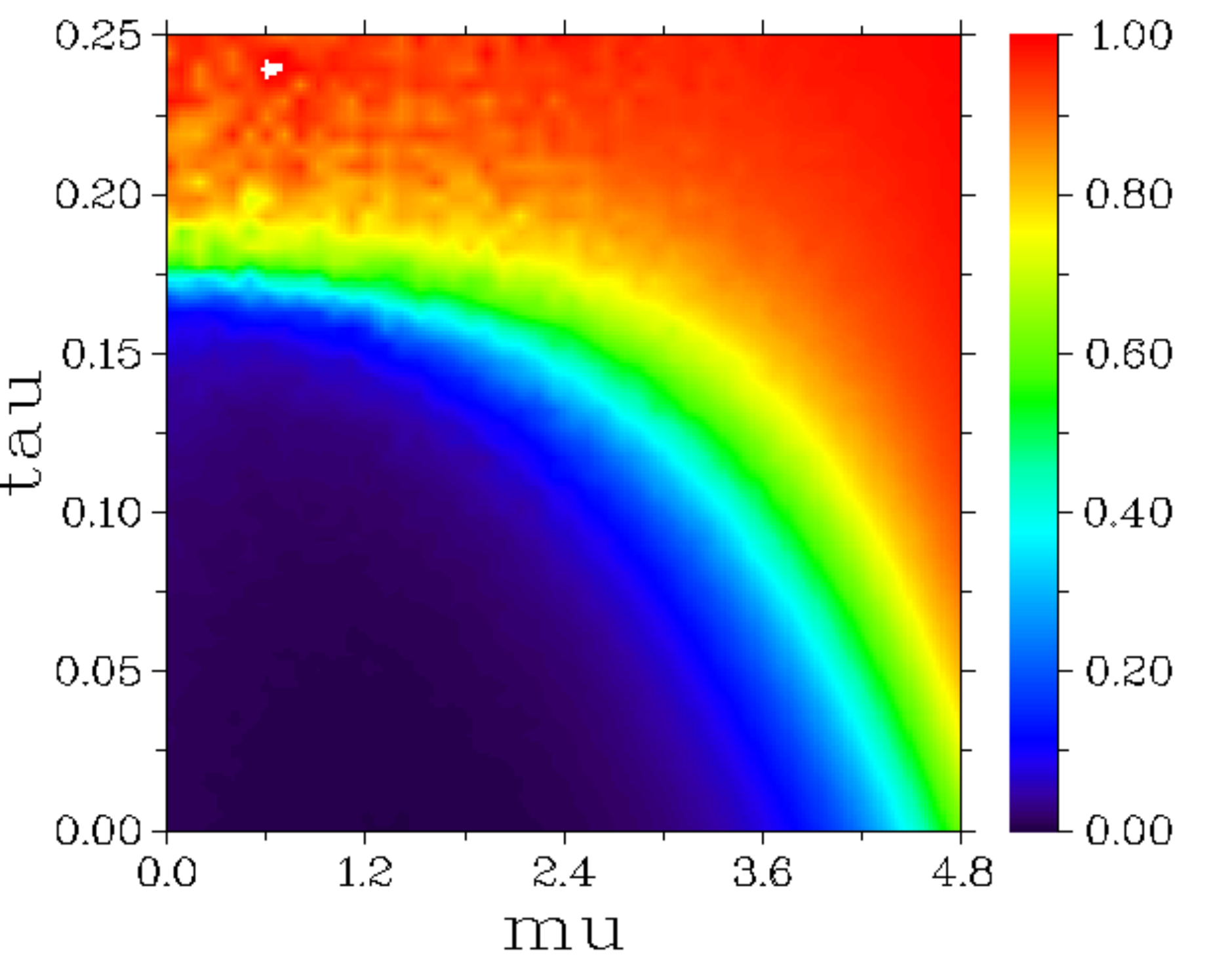} \hspace{0.1cm}
  \includegraphics[height=5.cm]{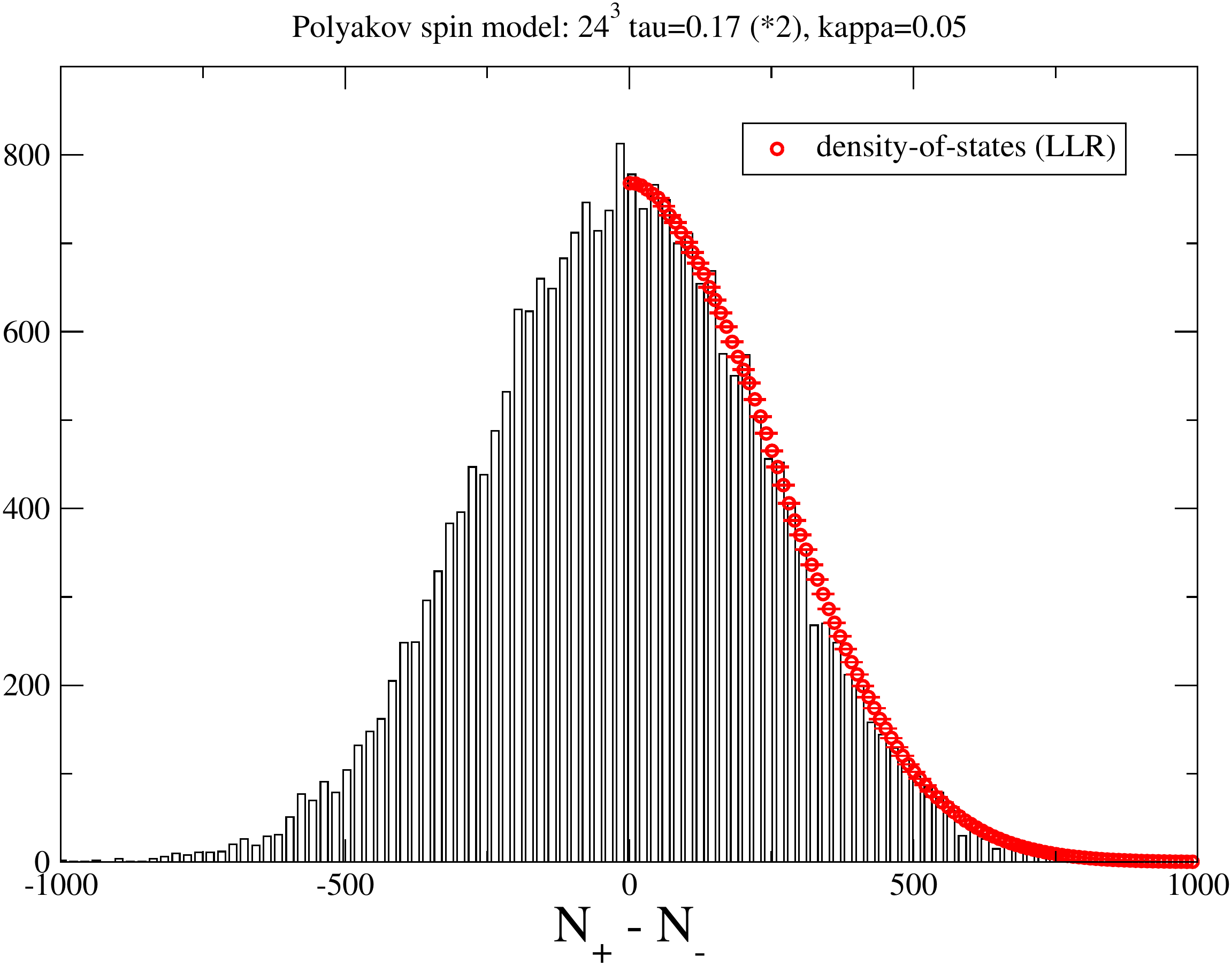}
}
\caption{Left: Phase diagram of the $Z_3$ theory as function of the
  chemical potential $\mu $ and $\tau $ (say temperature). 
  Right: The probability distribution of the imaginary part of the
  action. }
\label{fig:4}
\end{figure}
\begin{figure}[htb]
\centerline{%
  \includegraphics[height=6cm]{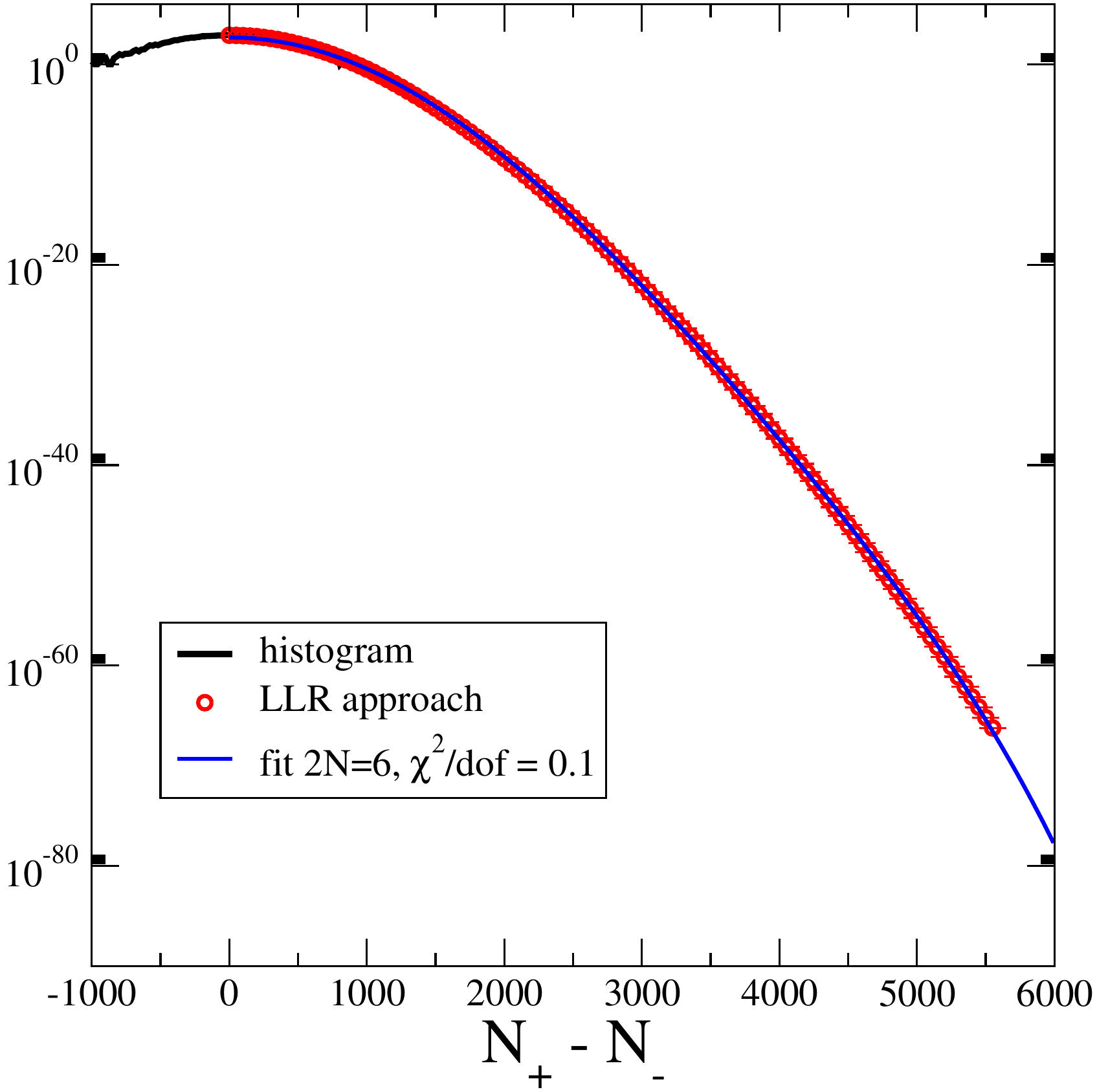} \hspace{0.1cm}
  \includegraphics[height=6cm]{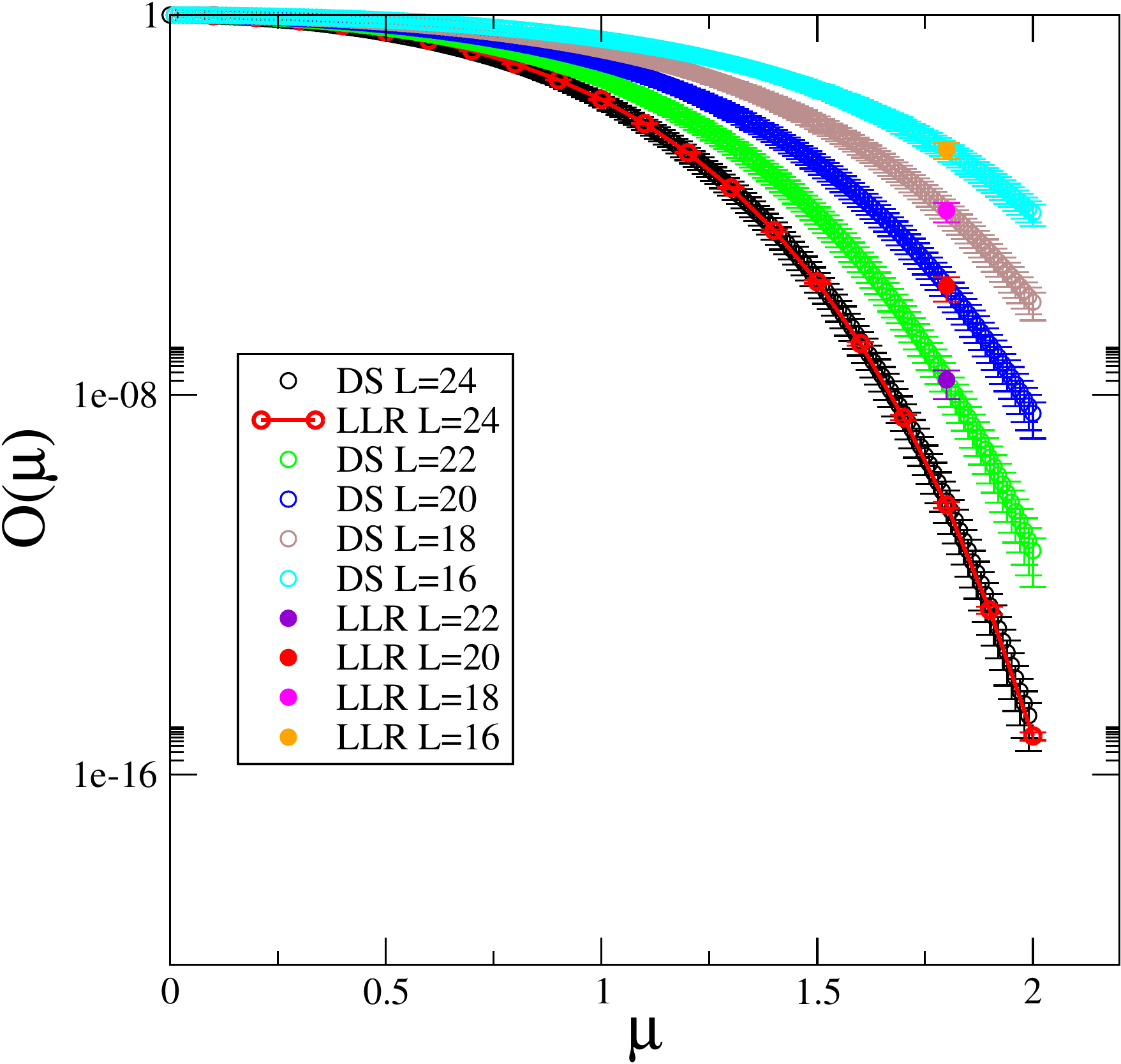} 
}
\caption{Left: Same as figure~\ref{fig:4}, right panel, on a
  logarithmic scale (result from~\cite{Gattringer:2016kco}). Right:
  LLR-result for the overlap factor $Q(\mu)$ as function of $\mu $
  (result from~\cite{Langfeld:2014nta}). 
}
\label{fig:5}
\end{figure}
The $Z_3$ theory in three dimensions is inspired by QCD if the 
$Z_3$ degrees of freedom on site are identified with the Polyakov line. 
The partition function as well as the action of the system are given by 
\bea
Z(\mu) &=& \sum _{\{\phi\}} \; \exp \Bigl( S[\phi] \Bigr) \; , \; \;
\; 
S[\phi] = \tau \sum _{x,\nu } \phi_x \, \phi^\ast _{x+\nu} + 
\sum_x \, \Bigl( \eta \phi_x + \bar{\eta } \phi^\ast _x \Bigr) \; , 
\label{eq:15} 
\ena 
with $ \eta = \kappa \, \mathrm{e}^{\mu } $ and $ \bar{\eta }  = \kappa \,
\mathrm{e}^{- \mu } $. For non-vanishing chemical potential, we have
$\eta \not= \bar{\eta }^\ast $ and the theory has a sign problem.  
Note, however, that this theory has a real dual
formulation~\cite{DeGrand:1983fk,Patel:1983sc,Mercado:2011ua}, and 
can be efficiently simulated with the flux algorithm\cite{Mercado:2012yf}. 
The phase diagram can be readily calculated (see figure~\ref{fig:4}, left
panel for our result) and bears a certain similarity of what we expect
for the QCD phase diagram. Here, it serves as an ideal benchmark 
for testing the generalised LLR approach~\cite{Langfeld:2014nta}.  

The probability distribution of the imaginary part $\rho _\beta (s)$
(see~(\ref{eq:8})) can be obtained by generating lattice configurations
with respect to the phase quenched theory and by subjecting the
imaginary part to a histogram. The result is shown in the right panel
of figure~\ref{fig:4}. Alternatively, we can calculate $\rho _\beta
(s)$ using the LLR formalism. The result is also shown in
figure~\ref{fig:4}: a good agreement of the LLR result with the
histogram is observed. We point out that the histogram method fails to
produce an accurate estimate for $\rho _\beta (s)$ at large imaginary
parts since hardly any events are recorded in this case. This is a
manifestation of the overlap problem. The LLR method solves this
overlap problem, and, by virtue of the exponential error suppression,
produces very good results over many orders of magnitude (see
figure~\ref{fig:5}, left panel). 

In order to obtain the overlap $Q(\mu)$, a Fourier transform of the
probability distribution $\rho _\beta (s)$ must be carried out.
Since the overlap exponentially decreases with the volume, the result
of the Fourier transform will generically produce a very small signal.
Despite of the precision for $\rho _\beta (s)$ that can be reached
with the LLR method, {\it compressing} the numerical data into an
analytic model with few parameters has proven
essential~\cite{Gattringer:2016kco}: 
\be 
\ln \rho (n) =  \lim _{N \to \infty } \sum _{k=1}^N c_k \; f_k(n)
\; , 
\label{eq:19}
\en 
where $f_k(n)$ are basis functions. The approximation arises from the
truncation of the above sum. For the $Z_3$ spin system,
good results are obtained by using powers of $n$: $f_k(n) =
n^{2k}$. Here we have exploited the symmetry $\rho (-n) = \rho(n)$,
which eliminates odd powers of $n$ from the basis. Figure~\ref{fig:5},
left panel, shows a typical result for such a fit. We find that for
$N$ bigger than some threshold $N_{th}$, the fits stabilise: the fit
results agree within error bars for $N \ge N_{th}$. Usually, the error
bars tend to increase for larger values of $N$. Hence, we use the fit
result for $N=N_{th}$ (and the corresponding statistical error bars
from the bootstrap analysis). Figure~\ref{fig:5}, right panel,
compares the LLR result with the results from a simulation of the dual
(real) theory. We find an excellent agreement despite of the fact that the
overlap becomes as small as $10^{-16}$ for the largest lattice size. 

\subsection{ QCD at finite densities of heavy quarks }

\begin{figure}[htb]
\centerline{%
  \includegraphics[height=6cm]{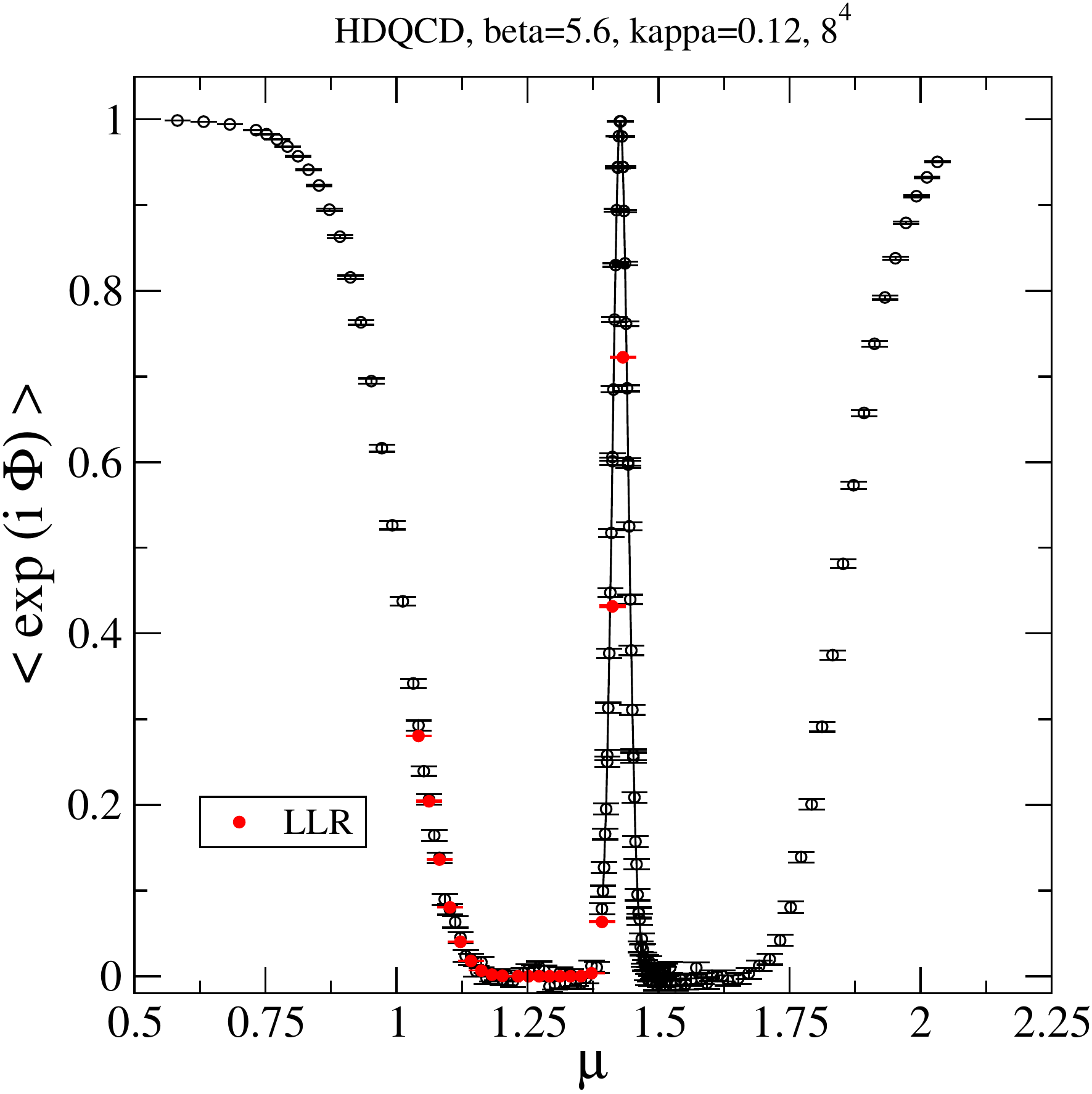} \hspace{0.1cm}
  \includegraphics[height=6cm]{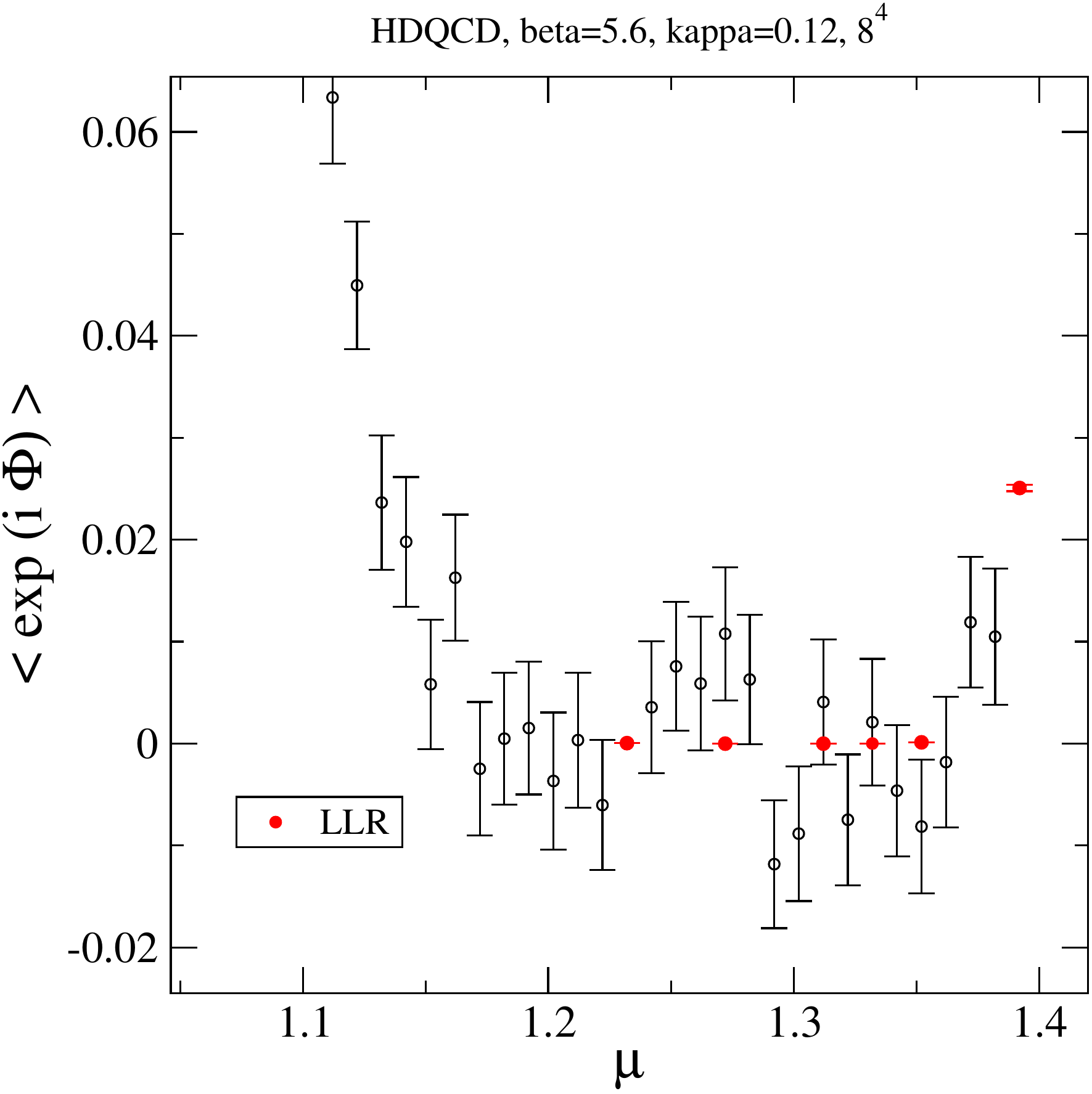} 
}
\caption{Left: The overlap factor for HDQCD as a function of the quark
  chemical potential (result from~\cite{Garron:2016noc}). Right:
  Detail of left panel. 
}
\label{fig:6}
\end{figure}
\begin{figure}[htb]
\centerline{%
  \includegraphics[height=6cm]{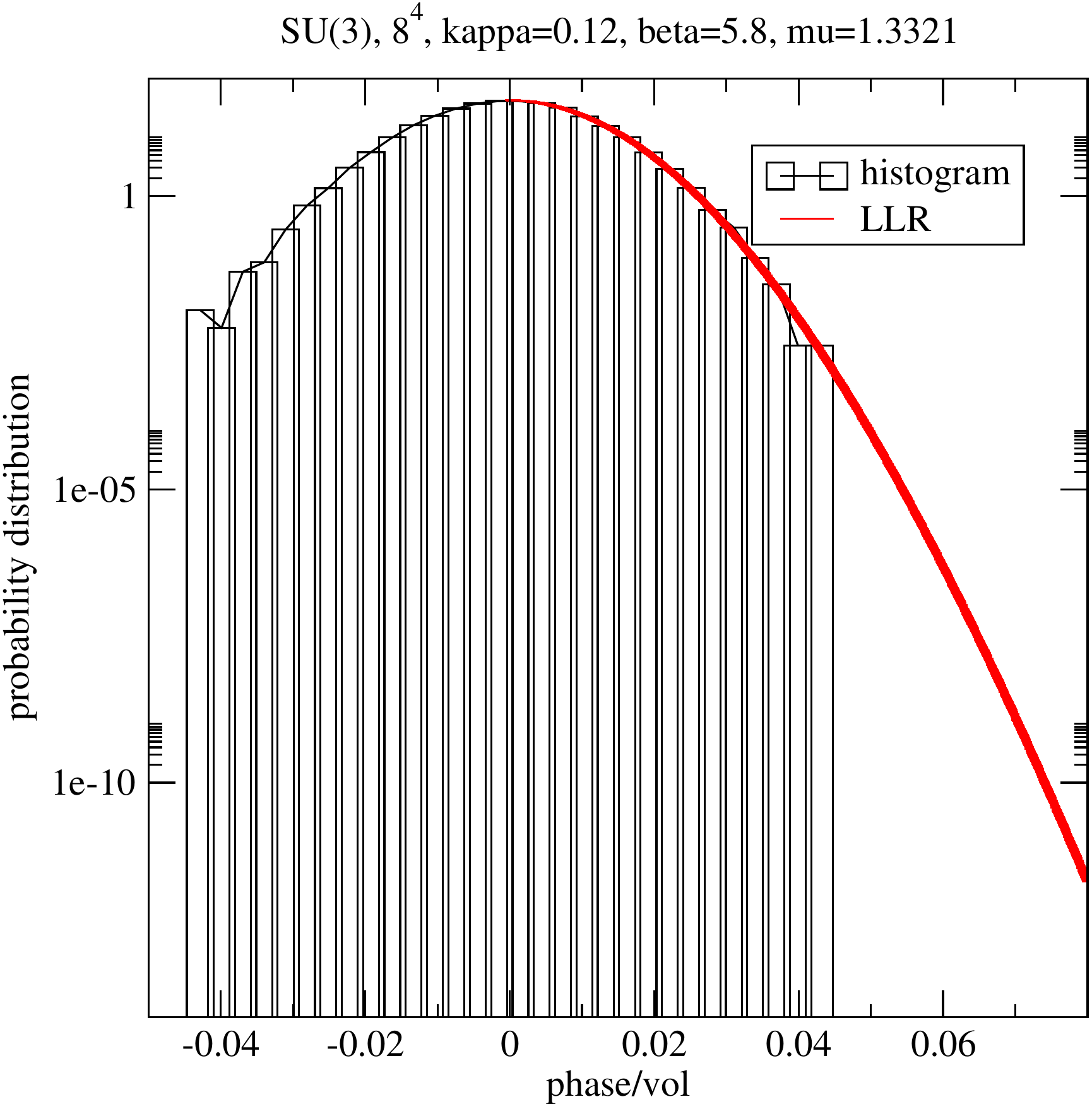} \hspace{0.1cm}
  \includegraphics[height=6cm]{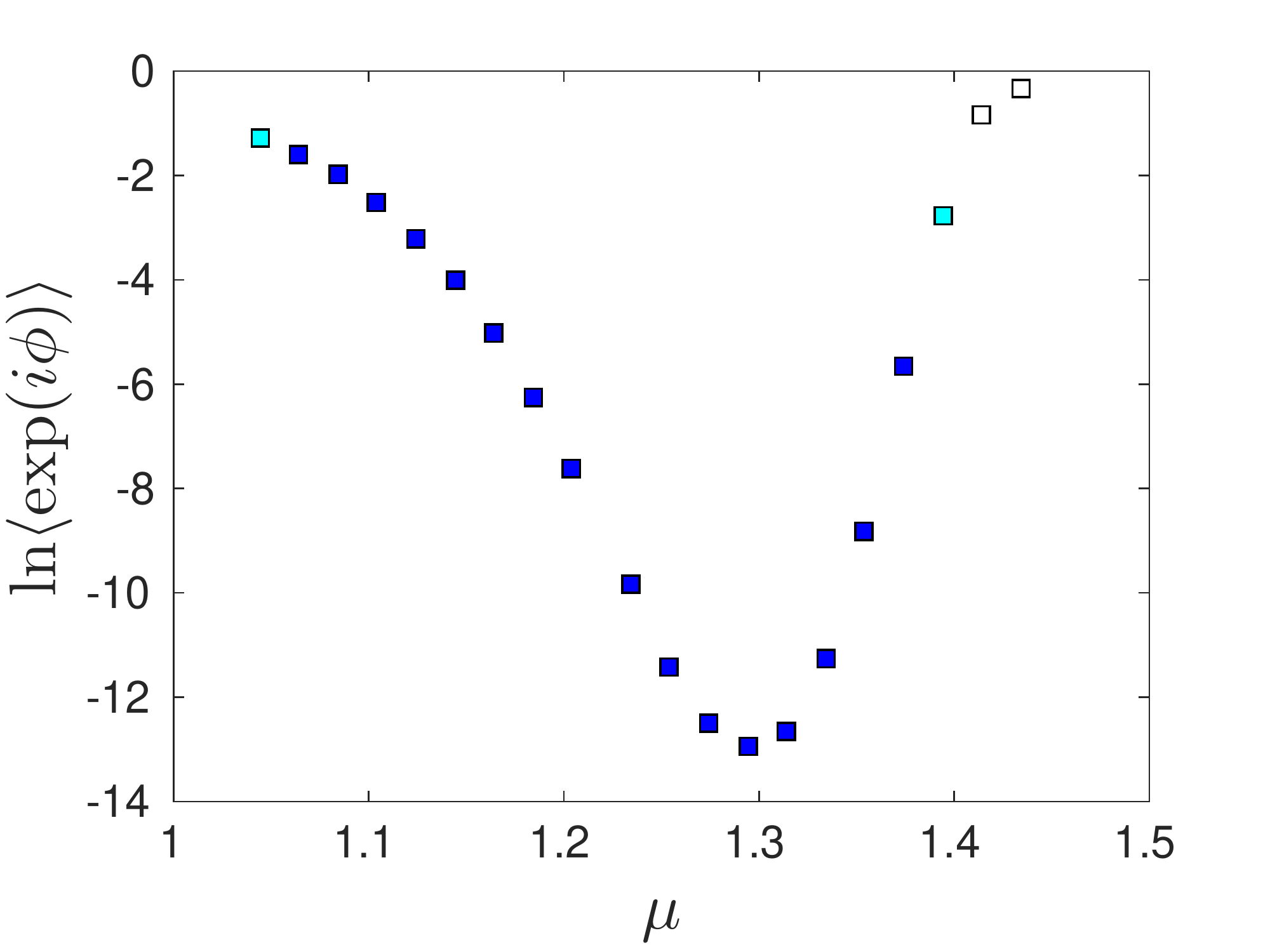} 
}
\caption{Left: Probability distribution of the imaginary part of the
  quark determinant. Right:   LLR result for the log of the overlap
  factor  (results from~\cite{Garron:2016noc}).
}
\label{fig:7}
\end{figure}
Our starting point is the QCD partition function
\be
Z(\mu) \; = \; \int {\cal D }U_\mu  \; \exp \{ \beta \, S_\mathrm{YM}[U] \}\;
\hbox{Det} M(\mu) \; ,
\label{eq:20}
\en
from which the quarks have been integrated out leaving us with the
quark determinant. In the so-called heavy-dense limit for large quark
mass $m$ and simultaneously large chemical potential $\mu $, the quark
determinant factorises 
into~\cite{Bender:1992gn,Blum:1995cb,Aarts:2014fsa,Aarts:2015yba,Rindlisbacher:2015pea}: 
\bea
\hbox{Det} \, M(\mu) &=& \prod _{\vec{x}} \; {\det} ^2 \Bigl( 1 \, + \,
\, \e ^{(\mu -m) / T} \, P(\vec{x}) \Bigr)
{\det } ^2 \Bigl( 1 \, + \, h
\, \e ^{ - (\mu+m) / T} \, P^\dagger (\vec{x}) \Bigr) \; ,
\nonumber 
\ena
where $m$ is the mass of the heavy quark, $T = 1/ N_t a $ is the temperature with
$a$ the lattice spacing and $N_t$ the number of lattice points in the 
temporal direction. At small temperatures $T \ll m$, we can ignore the
latter determinant in the latter equation. The theory is then not only real
at vanishing chemical potential, but also at threshold
$\mu=m$ (also called ``half-filling'') and the theory exhibits a
particle-hole
duality~\cite{Rindlisbacher:2015pea,Garron:2016noc}. Adopting a
standard re-weighting approach, we find for the overlap factor the
result shown in figure~\ref{fig:6}. At intermediate values for the
chemical potential, we do encounter a strong sign problem: the
re-weighting method produces results that are within statistical
errors compatible with zero implying that we have lost the signal in
the noise. 

To tackle this problem, we again employed the LLR approach to get high
quality results for the probability distribution of the phase of the
quark determinant. We checked that our results agree with those from a
straightforward histogram method (see figure~\ref{fig:7}). Based upon
our experience with the $Z_3$ theory, we adopted the same method to
obtain the density's Fourier transform (see (\ref{eq:19}) and
discussions below). Our final result for the overlap nicely agrees in
regions of the chemical potential where re-weighting can produce
statistical significant results (see figure~\ref{fig:6}). Note,
however, that the LLR result has error bars that are five orders of
magnitude smaller. Our final high quality result for the overlap is
shown in figure~\ref{fig:7}, left panel, and we refer the reader
to~\cite{Garron:2016noc} for further discussions. 

\medskip \noindent
{\bf Acknowledgements:} \hfill \break 
We thank  N.~Garron, A.~Rago and R.~ Pellegrini for helpful
discussions. We are grateful to the HPCC Plymouth and HPC Wales
for support received for carrying out numerical simulations.  KL is 
supported by  the  Leverhulme Trust (grant RPG-2014-118) and by
STFC (grant ST/L000350/1). BL is supported by STFC (grant ST/L000369/1).

{\small 
\bibliography{bk_lit}
}


\end{document}